\documentclass[aps,nofootinbib,preprintnumbers,citeautoscript]{revtex4}
\usepackage{amsmath,amssymb}
\usepackage{enumerate}
\usepackage{graphicx}
\usepackage{amsthm}
 \usepackage{mathtools}
 \usepackage{textcomp}
  \usepackage{amssymb}
  \usepackage{amsthm}
  \usepackage{wasysym}
  \usepackage{amsmath}
  \usepackage[export]{adjustbox}
 
\usepackage{standalone}
  \usepackage{url}
\date{\today}

\begin{document}
\title{McEliece Cryptosystem Based On Extended Golay Code}

\author{Amandeep Singh Bhatia$^\ast$ and  Ajay Kumar  \\
\textit{Department of Computer Science, Thapar university, India} \\
E-mail: $^\ast$amandeepbhatia.singh@gmail.com}

\begin{abstract}
With increasing advancements in technology, it is expected that the emergence of a quantum computer will potentially break many of the public-key cryptosystems currently in use.  It will negotiate the confidentiality and integrity of communications. In this regard, we have privacy protectors (i.e. Post-Quantum Cryptography), which resists attacks by quantum computers, deals with cryptosystems that run on conventional computers and are secure against attacks by quantum computers. The practice of code-based cryptography is a trade-off between security and efficiency. In this chapter, we have explored The  most successful McEliece cryptosystem, based on extended Golay code [24, 12, 8]. We have examined the implications of using an extended Golay code in place of usual Goppa code in McEliece cryptosystem. Further, we have implemented a McEliece cryptosystem based on extended Golay code using MATLAB. The extended Golay code has lots of practical applications. The main advantage of using extended Golay code is that it has codeword of length 24, a minimum Hamming distance of 8  allows us to detect 7-bit errors while correcting for 3 or fewer errors simultaneously and can be transmitted at high data rate.
\end{abstract}
\maketitle

%\newpage

%\tableofcontents

\theoremstyle{plain}

\newtheorem{thm}{Theorem}

\theoremstyle{definition}
\newtheorem{defn}{Definition}
\newtheorem{exmp}{Example}

\section{Introduction}
Over the last three decades, public key cryptosystems (Diffie-Hellman key exchange, the RSA cryptosystem, digital signature algorithm (DSA), and Elliptic curve cryptosystems) has become a crucial component of cyber security. In this regard, security depends on the difficulty of a definite number of theoretic problems (integer factorization or the discrete log problem). Table 1 represents the present status of several cryptosystems \cite{128}. Shor's algorithm is well-known in the field of cryptography given its potential application in cracking various cryptosystems, such as RSA algorithm and elliptic curve cryptography \cite{129}. These all public key cryptosystems can be attacked in polynomial time using Shor's algorithm. 

\begin{table} [h]
	\centering
	\caption{Impact of Quantum computing on cryptographic algorithms \cite{128}}
	\begin{tabular}{ |p{7.2cm}|p{3cm}| }
		\hline 
		\textbf{Cryptosystem} & \textbf{Broken by Quantum algorithms?}\\
		\hline
		Diffie-Hellman key-exchange \cite{153} &	Broken\\
		RSA public key encryption \cite{154} &	Broken\\
		Algebraically Homomorphic \cite{155}	& Broken\\
		Buchmann-Williams key-exchange \cite{157} &	Broken\\
		Elliptic curve cryptography \cite{156}	& Broken\\
		NTRU public key encryption \cite{159} &	Not broken yet\\
		McEliece public key encryption \cite{158} &	Not broken yet\\
		Lattice-based public key encryption \cite{160} &	Not broken yet\\
		\hline
	\end{tabular}
\end{table}

Post-Quantum Cryptography offers secure alternatives. The goal of post-quantum cryptography is to develop cryptographic systems that are secure against both quantum and classical computers, and compatible with existing communications protocols and networks. Apart from RSA, DSA, and ECDSA, there are other important classes of cryptographic systems which include Code-based, Lattice-based, Hash-based, Multivariate-quadratic-equations and Secret-key cryptosystem.

Code-based cryptography \cite{161} generally refers to cryptosystems in which the algorithmic primitive uses an error correcting code C. This primitive may consist of adding an error to a word of C or in computing a syndrome relatively to a parity check matrix of C. There are several codes for which efficient decoders are known. Fig 1 shows the several codes proposed and broken in code-based cryptography.

In 1949, Golay \cite{162} discovered Golay codes. A binary Golay code is a linear error-correcting code used in digital communication. Golay codes are perfect codes in which the Hamming spheres surrounding the codewords fill the Hamming space without overlap. These spheres have a radius \textit{e}, which can correct \textit{e} errors and their codewords separated from each other by a distance \textit{d}=\textit{e}+1. Perfect codes possess complete bounded-distance decoders and satisfy the Hamming bound with equality. If Golay codes are augmented with bit interleaving technique, it enables us to correct burst errors \cite{163}.

\section{Preliminaries}
In this section, some preliminaries and basic notations are given, which will be used throughout the chapter.
\begin{itemize}
	\item \textit{Linear code}: Linear code \textit{C} \cite{164} of length \textit{n} and dimension \textit{k} over a field \textit{F} is a \textit{k}-dimensional subspace of the vector space $F^{n}_q$ with \textit{q} elements, a set of \textit{n}-dimensional vectors can be referred to as a [\textit{n}, \textit{k}]  code and elements of bits such that \textit{F}=GF(2)=\{0,1\}. If the minimum Hamming distance of the code is \textit{d}, then the code is called a [\textit{n}, \textit{k}, \textit{d}] code.
	\item \textit{Hamming distance}:  A Hamming distance \cite{163} $d_H(x, y)$ is the number of positions in which two codewords \textit{(x, y)} differ. Let \textit{C} be a [\textit{n}, \textit{k}] linear code over $F^{n}_q$ and $x=(x_1, x_2,..., x_n), y=(y_1, y_2,..., y_n)$ are two code words. 
	\begin{equation}
	d_H(x, y)= \mid i: x_i \neq y_i, 1 \leq i \leq n \mid
	\end{equation}
	\item \textit{Hamming weight}: A Hamming weight \cite{163} $wt_H(x)$ is defined as the number of non-zero positions in the codeword \textit{x}. Let \textit{C} be a [\textit{n}, \textit{k}]  linear code over $F^{n}_q$  and $x=(x_1, x_2,..., x_n)$  is a code word, such that 
	\begin{equation}
	wt_H(x)= \mid i: x_i \neq 0, 1 \leq i \leq n \mid
	\end{equation}
	
	\item	\textit{Generator matrix}: A generator matrix \cite{163} for \textit{C} is a \textit{$k \times n$} matrix \textit{G} having the vectors of $V=(v_1, v_2,..., v_k)$  as rows, which forms a basis of \textit{C} such that
	\begin{equation}
	C=\{mG: m \in F^{n}_q \}, ~G= \begin{bmatrix}
	v_1 \\ v_2 \\ ... \\ v_k
	\end{bmatrix}
	\end{equation}
	The matrix \textit{G} generates the code as a linear map: for each message $m \in F^{n}_q $, we obtain 
	the corresponding code word \textit{mG}.
	\item \textit{Dual code}: Let \textit{C} be a [\textit{n}, \textit{k}]  linear code over $F_q^n$. The dual code [26] of \textit{C} is the set, such that $C^\perp=\{x \in F_q^n:x.y=0, \forall y \in C\}$ .
	\item \textit{Parity matrix}: A $(n-k) \times n$  generator matrix \textit{H} is called parity-check matrix \cite{163} for codeword \textit{C}, which is described by 
	\begin{equation}
	C=\{ m \in F^{n}_q: mH^T=0\},
	\end{equation}
\end{itemize}

\section{Prior work}
Originally, Golay codes \cite{162} were invented in the early 1950’s, and have experienced incredible responses in the last few years. In 1978, McEliece \cite{158} proposed an asymmetric encryption cryptosystem based on Goppa codes, which remains unbroken, even after 15 years of adaptation of its proposal security parameters \cite{165}. Niederreiter \cite{166} proposed a knapsack-type cryptosystem based on Reed-Solomon codes. Sidelnikov and Shestakov \cite{164} attacked the Niederreiter cryptosystem and proved that it is insecure using Reed-Solomon codes as well as Goppa codes.

Sidelnikov \cite{167} proposed a public-key cryptosystem based on binary Reed-Muller codes. It offered a high security with transmission rate close to 1, and complexity of encryption and decryption process is low. Minder and Shokrollahi \cite{168} attacked the Sidelnikov public-key cryptosystem which generates a private key from a known public key. It has been shown that running time of the attack is subexponential using low weight finding algorithms.

Janwa and Moreno \cite{169} proposed a McEliece public key cryptosystems based on Algebraic-Geometric Codes (AGC). It shows the various aspects of McEliece cryptosystem, based on the larger class of q-ary algebraic-geometric Goppa codes and listed some open problems for future improvements. Faure and Minder \cite{170} presented an algorithm based on algebraic geometry codes to recover the structure of algebraic geometry codes defined over a hyperelliptic code. In 2014, Couvreur et al. \cite{171} constructed a polynomial time algorithm attack against public key cryptosystems based on algebraic-geometric codes.

In 2000, Monico et al. \cite{172} showed an efficient way of using low-density parity check codes in McEliece cryptosystem. In 2007, Baldi et al. \cite{173} introduced a new variant of McEliece cryptosystem, based on quasi-cyclic low-density parity check (QCLDPC) codes. Furthermore, they examined the relevant attacks against LDPC and QCLDPC. Londahl and Johansson \cite{174} constructed a new version of McEliece cryptosystem based on convolutional codes. Landais and Tillich \cite{175} implemented an attack against McEliece cryptosystem based on convolutional codes. Various researchers proposed modified McEliece cryptosystems by replacing Goppa codes and using different error-correcting codes, e.g. algebraic geometric codes (AGC), low-density parity check codes (LDPCC) or convolutional codes. However, all of these schemes have proven to be insecure, making Goppa codes a standard solution.

\section{McEliece Cryptosystem}
McEliece cryptosystem is based on linear error-correcting code for creating public and private key. Binary Goppa code \cite{158} is used as the error-correcting code in McEliece cryptosystem. The secret key can be drawn from the various alternate codes. Several versions of McEliece cryptosystem were proposed using various secret codes such as Reed-Solomon codes, concatenated codes and Goppa codes. Interested researchers can study the original McEliece cryptosystem algorithm described in \cite{158}. 

\section{Golay Codes}
Golay codes can be classified into binary and ternary Golay codes. Furthermore, binary Golay codes are divided into extended ($G_{24}$) and perfect ($G_{23}$) binary Golay codes \cite{162, 163}. The extended binary Golay code $G_{24}$ is a [24, 12, 8] code, which encodes 12 bits of data into a word of 24-bit length in such a way that any 3-bit errors can be corrected or any 7-bit errors can be detected.

$$A_{23}=\begin{bmatrix}
0 & 1 & 1 & 1 & 1 & 1 & 1 & 1 & 1 & 1 & 1 \\
1 & 1 & 1 & 0 & 1 & 1 & 1 & 0 & 0 & 0 & 1 \\
1 & 1 & 0 & 1 & 1 & 1 & 0 & 0 & 0 & 1 & 0 \\
1 & 0 & 1 & 1 & 1 & 0 & 0 & 0 & 1 & 0 & 1 \\
1 & 1 & 1 & 1 & 0 & 0 & 0 & 1 & 0 & 1 & 1 \\
1 & 1 & 1 & 0 & 0 & 0 & 1 & 0 & 1 & 1 & 0 \\
1 & 1 & 0 & 0 & 0 & 1 & 0 & 1 & 1 & 0 & 1 \\
1 & 0 & 0 & 0 & 1 & 0 & 1 & 1 & 0 & 1 & 1 \\
1 & 0 & 0 & 1 & 0 & 1 & 1 & 0 & 1 & 1 & 1 \\
1 & 0 & 1 & 0 & 1 & 1 & 0 & 1 & 1 & 1 & 0 \\
1 & 1 & 0 & 1 & 1 & 0 & 1 & 1 & 1 & 0 & 0 \\
1 & 0 & 1 & 1 & 0 & 1 & 1 & 1 & 0 & 0 & 0 
\end{bmatrix}
$$

The perfect binary Golay code $G_{23}$ is a [23, 12, 7] code that is having a code word of length 23. It can be obtained from the extended binary Golay code by deleting one coordinate position. It is useful in the applications where a parity bit is added to each word for producing a half-rate code \cite{176}. It is constructed by a factorization $x^{23}-1$ over field $F_2^m$  such that: $x^{23}-1=(x-1)(x^{11}+x^9+x^7+x^6+x^5+x+1)(x^{11}+x^{10}+x^6+x^5+x^4+x^2+1)$, $g_1(x)= (x^{11}+x^9+x^7+x^6+x^5+x+1)$ and $g_2(x)= (x^{11}+x^{10}+x^6+x^5+x^4+x^2+1)$ are irreducible polynomials of degree (\textit{m}=11). These polynomials are reverse of each other and can generate the same cycle code words. Therefore, the generator matrix $12\times23$ of perfect binary Golay code is $G_{23}=[I_{12},A_{23}]$, where $I_{12}$ is $12\times12$ the identity matrix. Matrix $A_{23}$ is as follow:

\subsection{Binary extended Golay codes}
In 1977, extended Golay codes $G_{24}$ \cite{162} were used for error control on the Voyager 1 and 2 spacecraft launched towards Jupiter and Saturn. The perfect binary Golay code results into 3-byte extended Golay code by adding a parity bit. Some special properties of extended Golay Codes are:

\begin{itemize}
	\item $G_{24}$ is a self-dual code with a generator matrix $G=[I_{12} \mid A]$.
	\item	Parity check matrix for $G_{24}$  is $H=[A \mid I_{12}]$ \cite{177}.
	\item	Another generator and parity check matrix for $G_{24}$ are $G'=[A \mid I_{12}]$ and $H=G^t[\frac{A}{I_{12}}]$ respectively \cite{178}. 
	\item	The weight of every code word in $G_{24}$   is a multiple of 4 and distance is 8.
\end{itemize}
The extended Golay code generated by the $12 \times 24$ matrix $G=[I_{12} \mid A]$, where $I_{12}$ is $12 \times 12$     
the identity matrix and matrix \textit{A} is as shown below. 

$$A=\begin{bmatrix}
1 & 1 & 0 & 1 & 1 & 1 & 0 & 0 & 0 & 1 & 0 & 1 \\
1 & 0 & 1 & 1 & 1 & 0 & 0 & 0 & 1 & 0 & 1 & 1\\
0 & 1 & 1 & 1 & 0 & 0 & 0 & 1 & 0 & 1 & 1 & 1 \\
1 & 1 & 1 & 0 & 0 & 0 & 1 & 0 & 1 & 1 & 0 & 1 \\
1 & 1 & 0 & 0 & 0 & 1 & 0 & 1 & 1 & 0 & 1 & 1\\
1 & 0 & 0 & 0 & 1 & 0 & 1 & 1 & 0 & 1 & 1 & 1\\
0 & 0 & 0 & 1 & 0 & 1 & 1 & 0 & 1 & 1 & 1 & 1\\
0 & 0 & 1 & 0 & 1 & 1 & 0 & 1 & 1 & 1 & 0 & 1\\
0 & 1 & 0 & 1 & 1 & 0 & 1 & 1 & 1 & 0 & 0 & 1\\
1 & 0 & 1 & 1 & 0 & 1 & 1 & 1 & 0 & 0 & 0 & 1\\
0 & 1 & 1 & 0 & 1 & 1 & 1 & 0 & 0 & 0 & 1 & 1\\
1 & 1 & 1 & 1 & 1 & 1 & 1 & 1 & 1 & 1 & 1 & 0
\end{bmatrix}$$

\section{McEliece Cryptosystem using extended Golay code}
McEliece cryptosystem based on extended Golay code works similarly as McEliece cryptosystem, but it generates the secret matrix  \textit{G} with a different way, and different decoding procedure will be used for the decoding process. Golay code matrix \textit{A}  is having a cyclic structure, in which the second row is obtained by moving the first component to the last position. Similarly, each row of the matrix \textit{A}   can be obtained by a right shift of the previous row, except last one row. The matrix \textit{A} is being a part of both the generator and the parity check matrices of extended Golay code; its decoding procedure is very simple. The main idea is to replace the Goppa code used in McEliece by an extended Golay code that can be efficiently decoded.

\begin{figure}[!h]
	\centering
	\includegraphics[scale=0.4, frame]{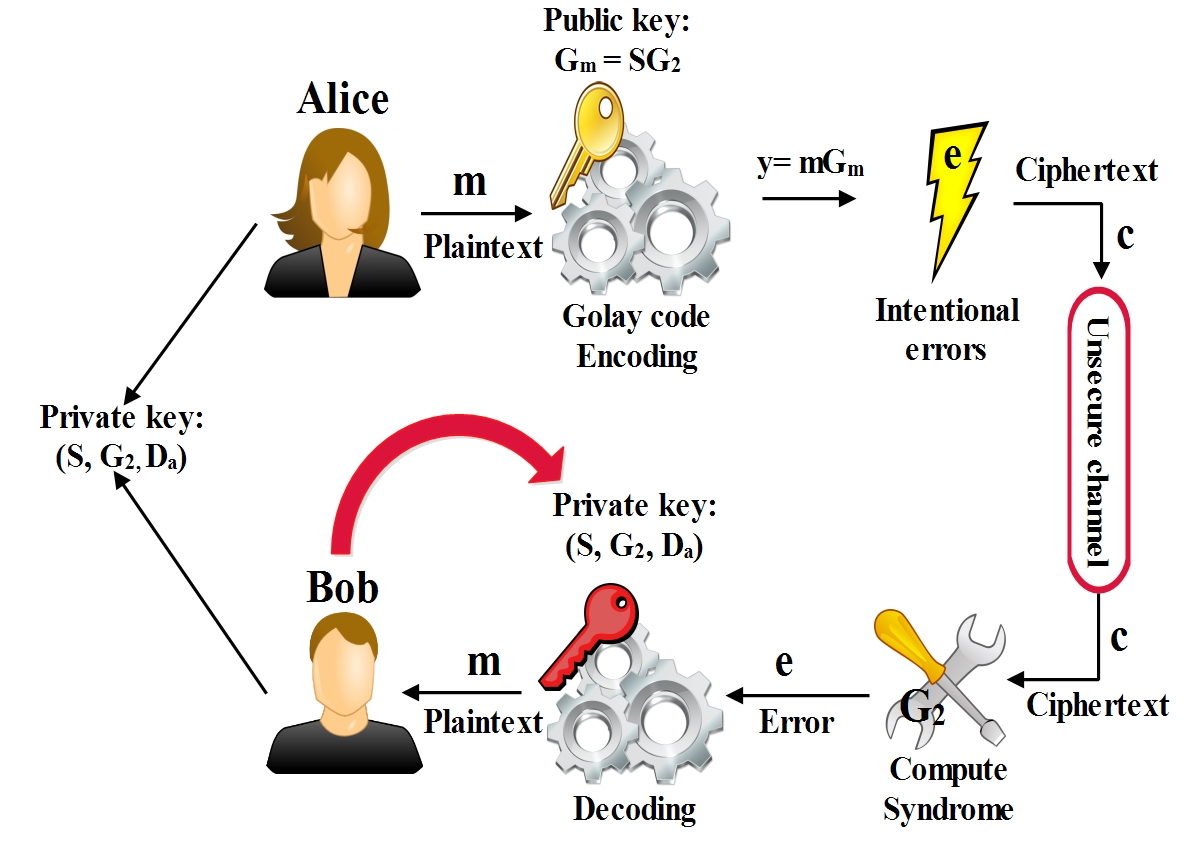}
	\caption{McEliece Cryptosystem using extended Golay code}
\end{figure}

\subsection{Key generation}
McEliece cryptosystem based on extended Golay code, $G_{24}$ encode 12-bits of data in  24-bit length of the word. Random permutation matrix (\textit{P}) acts on generator matrix (\textit{G}). Then, reorder the computed matrix and named it as $G_2$. Compute $G_m \leftarrow SG_2$  by the random invertible matrix (\textit{S}) and makes the public key ($G_m, t$)  and secret key ($S, G_2, D_a$)  correcting any 3-bit of errors. Key generation is described in algorithm 1. The detailed algorithm of McEliece cryptosystem based on extended Golay code is given below.

\begin{table}
	\centering
	\begin{tabular}{|p{14cm}|} 
		\hline
		\textbf{Algorithm 1: Key generation}\\
		\textbf{System parameters:} Let \textit{F} be a family of \textit{t}-error correcting  ($t \in N$)  \textit{q}-ary linear [\textit{n}, \textit{k}, \textit{d}] codes, where $t \ll n$.\\
		\textbf{Input:}  $G_{24}$ [24, 12, 8]  is an extended Golay which encodes (\textit{k}=12)  bits of data in a word of (\textit{n}=24) bit length and any (\textit{t}=3) bit errors can be corrected. \\
		\textbf{Output matrices:} \begin{itemize}
			\item 	Generate generator matrix \textit{G}: $k \times n$ generator matrix for code \textit{C} capable of correcting \textit{e} errors over \textit{F} of dimension \textit{k}. $G \leftarrow [I_{12} \mid A]$, where $I_{12}$ is $12 \times 12$ the identity matrix.
			\item	Generate permutation matrix \textit{P}: $n \times n$ is a random permutation matrix, having exactly 1 in every row and column; with all other entries is zero.
			\item	Compute $k \times n$  matrix $G_1=GP$, arrange $G_1$  in systematic format of generator matrix and named it as $G_2$.
			\item Generate a non-singular invertible matrix $S \in F_2^{k \times k}$ .
			\item	Compute $k \times n$   matrix $G_m \leftarrow SG_2$. 
			\item	Return public key: ($G_m, t$), private key: ($S, G_2, D_a$), where $D_a$ is an efficient decoding algorithm.
		\end{itemize} \\
		\hline
	\end{tabular}
\end{table}

\subsection{Encoding}
In encoding, the plaintext is a random non-zero binary vector of length \textit{k}, i.e. ($m \in F_2^k$). A ciphertext ($c \in F_2^n$) is the code word of the code with generator matrix  $G_m$ and we choose random error vector ($e \in F_2^n$)  exactly of weight \textit{t}. The encoding process is defined in the algorithm 2.

\begin{table} [!h]
	\centering
	\begin{tabular}{|p{14cm}|} 
		\hline
		\textbf{Algorithm 2: Encoding}\\
		\textbf{Input:}  Public key ($G_m \in F_2^{k \times n}$), message ($m \in F_2^k$), error vector ($e \in F_2^n$).   \\
		\textbf{Output:} Ciphertext ($c \in F_2^n$) \\
		\hline
		Compute $y \leftarrow m G_m$ \\
		Add error vector $c \leftarrow y+e$\\
		Return \textit{c} \\
		\hline
	\end{tabular}
\end{table}
\subsection{Decoding}
The decoding process is defined in the algorithm 3. It uses the decoding procedure of extended Golay code, whereas original McEliece cryptosystem uses Patterson's algorithm for the decoding process. 

\begin{table} [!h]
	\centering
	\begin{tabular}{|p{14cm}|} 
		\hline
		\textbf{Algorithm 3: Decoding}\\
		\textbf{Input:}  Ciphertext ($c \in F_2^n$), Private key: ($S, G_2, D_a$)    \\
		\textbf{Output:} Original message ($m \in F_2^k$) \\
		\hline
		Compute the encoded message $y_1 \leftarrow c+e$, where \textit{e} is calculated by calling subroutine $D_a(c, G_2)$. \\
		$y_1 \leftarrow mSG_2+ e$, compute message \textit{mS} by row reducing [$G_2^t \mid (mSG_2)^t$]. \\
		Multiply \textit{mS} by $S^{-1}$.\\
		Return \textit{m} \\
		\hline
	\end{tabular}
\end{table}

Here, we call a subroutine $D_a(c)$, which computes an error vector described in the algorithm 4. Therefore, on reading input a ciphertext ($c \in F_2^n$), it generates an output as the original message ($m \in F_2^k$). In step 1, it computes a syndrome using private key $G_2$ checks whether the weight of syndrome $s_1$ is less than or equal to 3. If yes, then it returns an error vector \textit{e}=[$s_1,000000000000$]. Otherwise, it checks the weight of ($s_1+A_i$) is less than or equal to 2, then the error vector is  \textit{e}=[$s_1+A_i,j_i$]. If it does not satisfy the first condition, then further it computes the second syndrome $s_2$ and checks whether the weight of syndrome $s_2$ is less than or equal to 3. If yes, then it returns an error vector \textit{e}=[$000000000000,s_2$]. Otherwise, it checks the weight of ($s_2+A_i$) is less than or equal to 2, then the error vector is \textit{e}=[$j_i, s_2+A_i$]. In any case, if both the conditions do not satisfy and the error pattern \textit{e} is not yet determined, then it requests retransmission. Finally, \textit{mS} is found by row reducing form and the original message is computed by multiplying \textit{mS}  by $S^{-1}$.

\begin{table} [!h]
	\centering
	\begin{tabular}{|p{14cm}|} 
		\hline
		\textbf{Algorithm 4: $D_A(c, G_2$)}\\
		\textbf{Input:}  Ciphertext ($c \in F_2^n$), generator matrix (private): ($G_2$)    \\
		\textbf{Output:} Error vector ($e \in F_2^n$) \\
		\hline
		Compute the first syndrome: $s_1 \leftarrow cG_2$ \\
		\textbf{If} $wt(s_1) \leq 3$, then\\ 
		~ ~ ~ ~ Return \textit{e} $\leftarrow$ [$s_1,000000000000$]  \\
		\textbf{Else If} $wt(s_1+A_i) \leq 2$, then \\      
		~ ~ ~ ~ Return \textit{e} $\leftarrow$ [$s_1+A_i,j_i$], where $j_i$ the word of length 12 with 1  in the $i^{th}$\\ ~ ~ ~ ~ position and 0 elsewhere in  $I_{12}$ identity matrix.  \\
		\textbf{Else}\\
		~ ~ ~ ~ Compute the second syndrome: $s_2 \leftarrow s_1A$\\
		~ ~ ~ ~ \textbf{If} $wt(s_2) \leq 3$, then \\
		~ ~ ~ ~ ~ ~ ~ ~ Return \textit{e} $\leftarrow$ [$000000000000, s_2$] \\
		~ ~ ~ ~ \textbf{Else If} $wt(s_2+A_i) \leq 2$, then \\
		~ ~ ~ ~ ~ ~ ~ ~Return \textit{e} $\leftarrow$ [$j_i, s_2+A_i$] \\
		~ ~ ~ ~ ~ ~ ~ ~\textbf{Else If} the error pattern \textit{e} is not yet determined, then request\\ ~ ~ ~ ~ ~ ~ ~ ~retransmission.  \\
		\hline
	\end{tabular}
\end{table}

\subsection{Security}
The security of the proposed McEliece cryptosystem depends on the difficulty level to decode \textit{y}  into message \textit{m}. The attacker will have a tough time trying to separate $G_2$ from $G_m$  because he/she does not know \textit{P} and inverse of a matrix \textit{S}, which are not publicly available. Therefore, an attacker cannot find an error because it’s hard to recover the specific structure of the matrix $G_2$. Maximum-likelihood decoding can be used to recover error but making tables for big codes ($2^{n-k}=2^{24-12}=4096$) coset leader is a time-consuming and inefficient. It also needs more storage space and decoding time can be quite long also. Therefore, we rely on syndrome decoding of extended Golay code.

\section{Implementation of McEliece Cryptosystem based on Extended Golay Code}
We have used a personal computer to implement McEliece cryptosystem based on extended Golay code with the following specification: CPU Intel Core i3-3217U 1.80 GHz, RAM 2.00 GB, OS Windows 8 Enterprise 32 bit and MATLAB 7.11.0 (R2010b).

We have used $12 \times 24$ generator matrix \textit{G}=[$I_{12} \mid A$] to generate extended Golay code $G_{24}$, where $I_{12}$ is $12 \times 12$  the identity matrix. Fig. 2 shows the matrix \textit{A}, which is obtained by adding a parity bit at the end of each codeword of perfect Golay code $G_{23}$. We have used the random permutation matrix $24 \times 24$  to compute $G_1 \leftarrow GP$ as shown in Fig 3.

\begin{figure}[!h]
	\centering
	\includegraphics[scale=0.5, frame]{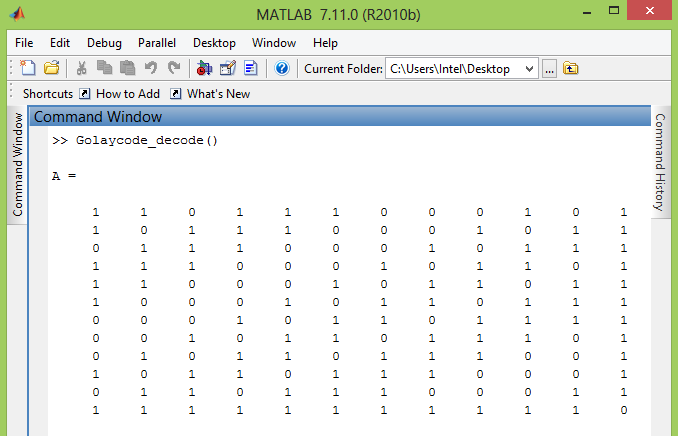}
	\caption{Generator polynomial matrix  \textit{A} of $G_{24}$}
\end{figure}

\begin{figure}[!h]
	\centering
	\includegraphics[scale=0.38, frame]{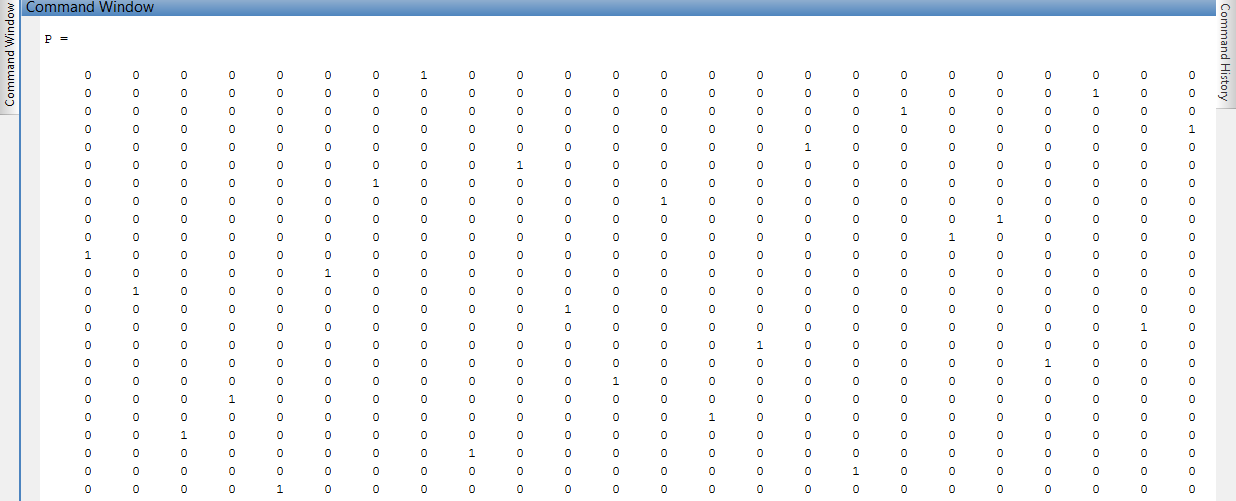}
	\caption{Random permutation matrix \textit{P}}
\end{figure}

We have used the random function to generate $12 \times 12$ a random invertible matrix \textit{S} of binary numbers. The matrix $G_1$ is reordered and renamed as $G_2$, then we computed $G_m=SG_2$, where $G_m$ is the encoding matrix. Fig. 4 represents the random invertible matrix \textit{S}.
Furthermore, $G_m$ encoding matrix results in public key: $(G_m, t)$, the private key consists of a random matrix \textit{S}, systematic generator matrix  $G_2$ and $D_a$ efficient decoding algorithm such that ($S, G_2, D_a$). We have used random plaintext \textit{m} of length 12 and random error vector \textit{e} of length is 24 having weight ($wt \leq 3$). Then, we compute codeword by $y=mG_m$  and encode it by computing ciphertext  such that  $c=y+e$. Fig. 5 shows the computed $G_m$ matrix codeword, random error, and ciphertext.

\begin{figure}[!h]
	\centering
	\includegraphics[scale=0.45, frame]{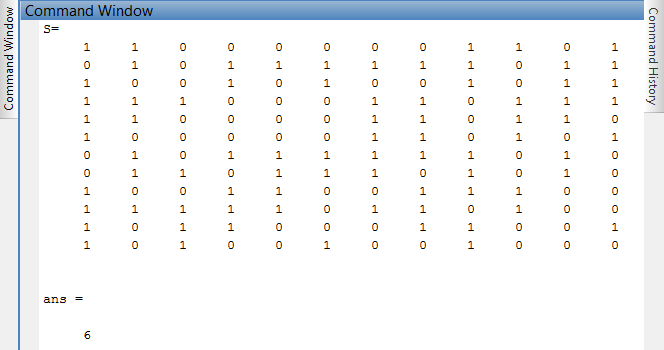}
	\caption{Random invertible matrix  \textit{S}}
\end{figure}

\begin{figure}[!h]
	\centering
	\includegraphics[scale=0.40, frame]{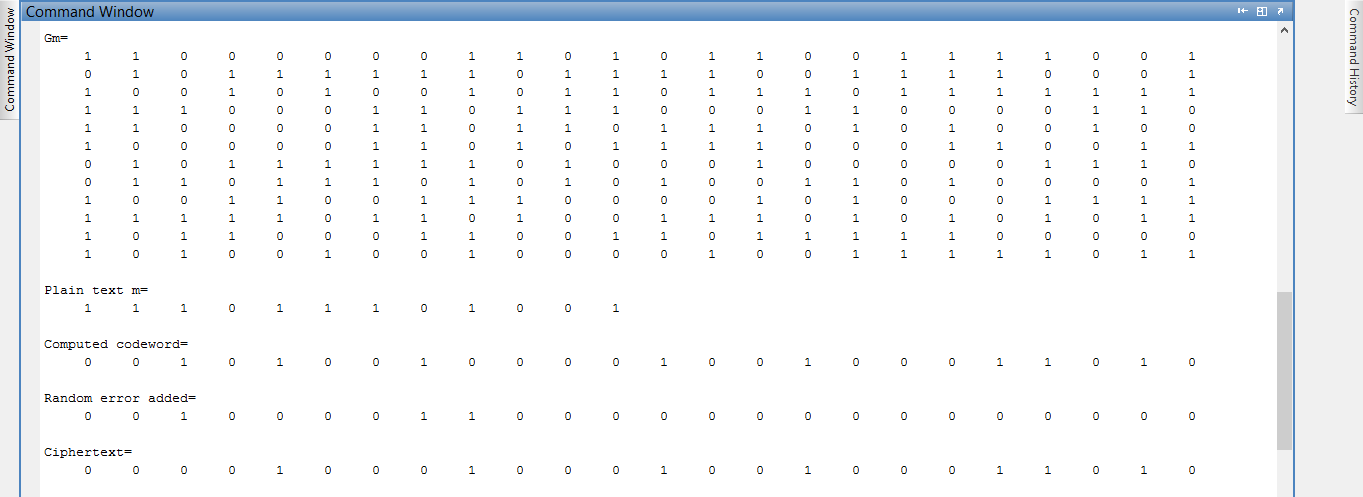}
	\caption{Generate ciphertext by adding intended error}
\end{figure}

During decoding, we call a subroutine as described in Algorithm 4 for computing an error \textit{e} by using private key $G_2$. Further, we recovered the actual codeword such that $y=c+e$. Fig 6 shows the calculated syndrome for error detection in the ciphertext.

\begin{figure}[!h]
	\centering
	\includegraphics[scale=0.4, frame]{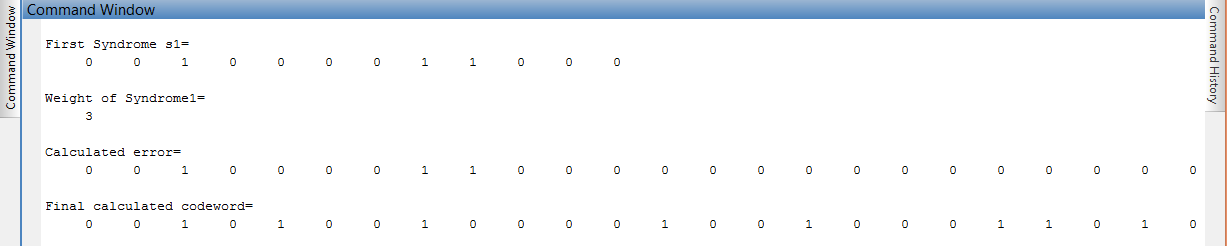}
	\caption{Error detection by calculating syndrome}
\end{figure}

Compute the error and actual codeword; we recover the plaintext by multiplying it with the inverse of \textit{S}. Fig. 7 shows the actual message sent over the channel.

\begin{figure}[!h]
	\centering
	\includegraphics[scale=0.5, frame]{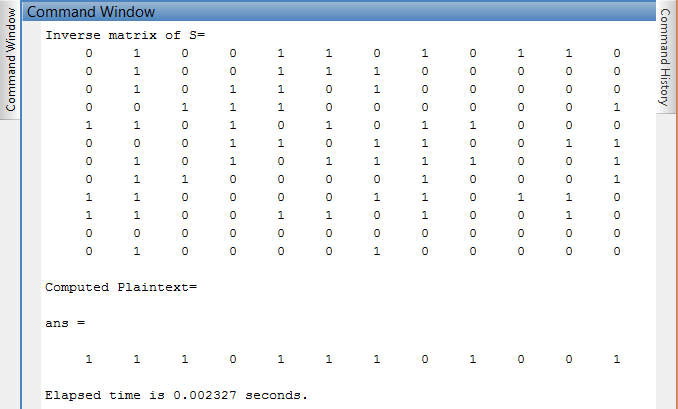}
	\caption{Decoding of Ciphertext}
\end{figure}

We have examined the McEliece cryptosystem using extended Golay code. The developed system is effective and secure until \textit{S} is chosen sparse random matrix. It corrects up to three-bit errors per codeword. Sparse matrices make it efficient and it allows a significant compression. Moreover, we have implemented the McEliece cryptosystem using extended Golay code and designed a finite state machine for its decoding component. In future, we will design McEliece cryptosystem using extended Golay code associated with bit interleaving technique to correct bursts of errors per codeword. 

\section{Conclusion}

  In this paper, we have examined the McEliece cryptosystem using extended Golay code. The
  developed system is effective and secure until S is chosen sparse random matrix. It corrects
  up to three-bit errors per codeword. Sparse matrices make it efficient and allows a
  significant compression. Moreover, we have implemented the proposed McEliece cryptosystem  using MATLAB. In future, we will design of McEliece cryptosystem using extended Golay code associated with bit
  interleaving technique to correct bursts of errors per codeword.

\section*{Acknowledgments}
Amandeep Singh Bhatia was supported by Maulana Azad National Fellowship (MANF), funded by Ministry of Minority Affairs, Government of India. 

\label{sec:test}
\bibliographystyle{elsarticle-num}
\bibliography{sample}
\end{document}